# Critical behavior of the insulator-to-metal transition in Te-hyperdoped Si

Mao Wang[1,*], A. Debernardi[2,‡], Wenxu Zhang,[3] Chi Xu[1,4], Ye Yuan[1,5], Yufang Xie[1,6], Y. Berencén[1], S. Prucnal[1], M. Helm[1,6] and Shengqiang Zhou[1]

[1]Helmholtz-Zentrum Dresden-Rossendorf, Institute of Ion Beam Physics and Materials Research, Bautzner Landstr. 400, 01328 Dresden, Germany

[2]CNR-IMM, sede Agrate Brianza, via Olivetti 2, I-20864, Agrate Brianza, Italy

[3]University of Electronic Science and Technology of China, State Key Laboratory of Electronic Thin Films and Integrated Devices, 610054 Chengdu, China

[4]Institute for Integrative Nanosciences, Institute for Solid State and Materials Research (IFW Dresden), Helmholtzstrasse 20, 01069 Dresden, Germany

[5]Songshan Lake Material laboratory, Dongguan, Guangdong, 523808, People's Republic of China

[6]Technische Universität Dresden, 01062 Dresden, Germany

**Abstract**

Hyperdoping Si with chalcogens is a topic of great interest due to the strong sub-bandgap absorption exhibited by the resulting material, which can be exploited to develop broadband room-temperature infrared photodetectors using fully Si-compatible technology. Here, we report on the critical behavior of the impurity-driven insulator-to-metal transition in Te-hyperdoped Si layers fabricated via ion implantation followed by nanosecond pulsed-laser melting. Electrical transport measurements reveal an insulator-to-metal transition, which is also confirmed and understood by density functional theory calculations. We demonstrate that the metallic phase is governed by a power law dependence of the conductivity at temperatures below 25 K, whereas the conductivity in the insulating phase is well described by a variable-range hopping mechanism with a Coulomb gap at temperatures in the range of 2-50 K. These results show that the electron wave-function in the vicinity of the transition is strongly affected by the disorder and the electron-electron interaction.

[*]Corresponding authors, email: m.wang@hzdr.de;

[‡]Author to whom correspondence should be addressed for first principles calculations, e-mail: alberto.debernardi@mdm.imm.cnr.it

# I. INTRODUCTION

The insulator-to-metal transition (IMT) in doped semiconductors is a prototypical example for a quantum phase transition and has been explored in many different systems, to large fraction in Si doped with shallow donors or acceptors. Generally, the IMT may be controlled by an external parameter $x$ which is experimentally accessible by impurity concentration ($N$), electric ($E$) or magnetic ($B$) field, or uniaxial stress ($S$) [1-6]. In current understanding, the IMT is driven by both disorder and interaction, thus being a mixed Anderson-Mott-Hubbard type of transition. In the simple model introduced by Mott [7], a criterion is derived relating the effective Bohr radius ($a_H$) of an isolated impurity with the critical density of impurities ($n_{crit}$) at the transition, given by $a_H n_{crit}^{1/3} \sim 0.25$. Experimentally, the impurity-mediated IMT has been studied extensively for shallow-level impurities (such as B, P, As and Sb) in heavily doped Si [5,8,9], where the critical transition concentration is below the solubility limit [10,11] and in the order of $10^{18}$ cm$^{-3}$ [12].

There are very few and relatively recent experimental studies of the IMT with deep-level impurities, such as chalcogens (S, Se, and Te) in Si [13-16]. According to Mott's theory the IMT is expected to occur at much higher concentrations ($n_{crit} \gg 10^{18}$ cm$^{-3}$) than for shallow donors, since electrons are much more tightly bound with a significantly reduced radius. In previous work chalcogen-hyperdoped Si with non-equilibrium concentrations was prepared using ion implantation followed by pulsed laser melting (PLM) [13,14,17]. In those studies, the transition from insulating to metallic conduction was identified with impurity concentrations exceeding $10^{20}$ cm$^{-3}$, which is four orders of magnitude larger than their equilibrium solubility limit of about $10^{16}$ cm$^{-3}$ [11]. Moreover, the nature of the IMT was explored by both experimental and computational approaches, where the IMT was owing to the delocalization of donor electrons above a critical donor concentration ($n_{crit}$) [1], which results in the formation of an intermediate band (IB) [14,18] and the merging of the broadened IB with the conduction band. A more systematic study on the conduction mechanism in the insulating and metallic phases and in the critical regime of the transition is however still lacking.

In the present work, we employ experimental and computational methods to identify the impurity-induced IMT in Si hyperdoped with Te and explore the critical behavior near the transition. We analyze the temperature-dependent conductivity of the Te-hyperdoped Si samples, which cover both sides of the insulator-to-metal transition. We find a power law dependence in metallic samples and a variable range hopping mechanism with a Coulomb gap in insulating samples. Combining with the first principles calculations, this work provides a consistent picture about the critical Te concentration and the conductivity behavior near the IMT.

## II. METHODS

### A. Computational details

First principles calculations of electronic structure were performed by plane-wave pseudo-potential techniques within the framework of Density Functional Theory as implemented in QUANTUM ESPRESSO (QE) open-source code [19]. The simulation of hyperdoped silicon was performed by the super-cell method by using ultrasoft pseudopotentials [20-22] in the separable form introduced by Kleinmann and Bylander [23], generated with a Perdew-Burke-Ernzerhof (PBE) exchange correlation functional. For substitutional single Te ($Te_{Si}$) and substitutional Te dimer ($Te_{Si}$-$Te_{Si}$), after structural relaxation we computed the electronic band-structure and the density of states (DOS) by solving the Kohn-Sham equations. All other computational parameters are the same as in Ref. [24] to which the interested reader can refer for further computational details. The Te concentration for both $Te_{Si}$ and $Te_{Si}$-$Te_{Si}$ refers to the percent concentration of Te atoms with respect to the amount of Si atoms. At a given concentration, to simulate $Te_{Si}$ or $Te_{Si}$-$Te_{Si}$, the size of the supercell (i.e. the total number of atomic sites) and/or the number of Te atoms in the super-cell are arranged accordingly, to obtain the stated Te concentration.

### B. Experimental details

Single-side polished Si (100) wafers (*intrinsic*, $\rho \geq 10^4$ $\Omega$·cm) were implanted with Te ions with six different fluences (as shown in Table 1) at room temperature. All Te concentrations were firstly calculated using the SRIM code [25] and then verified by Rutherford backscattering spectrometry (RBS) measurements (labeled in FIG. 1 and listed in table I). A combined implantation at energies of 150 keV and 50 keV with a fluence ratio of 2.5:1 was applied for a relatively uniform distribution of Te in the implantation region. Subsequently, ion-implanted samples were annealed using a pulsed XeCl excimer laser (Coherent COMPexPRO201, wavelength 308 nm, pulse duration 28 ns) in ambient air. A single laser pulse with an energy density of 1.2 J/cm$^2$ was used for processing samples presented in this work. After PLM, the damaged layer is recovered and reveals comparable crystalline quality as a virgin Si wafer [17]. Single-crystalline epitaxial regrowth of the Te-hyperdoped Si has also been confirmed by high resolution transmission electron microscopy. Moreover, Te is found to be uniformly distributed within the top 120 nm of the Si wafer without the formation of extended defects, secondary phases nor cellular breakdown [17]. During the annealing process, the whole amorphous implanted region was molten and then recrystallized with a solidification speed in the order of 10 m/s while cooling down [26]. This condition allows for Te concentrations beyond the solid solubility limit of Te in Si while preserving the epitaxial single-crystal growth.

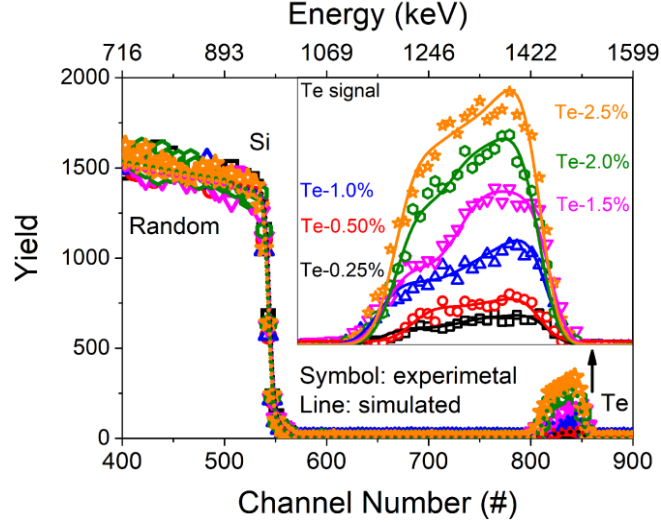

FIG. 1. RBS random spectra (symbols) and the fit (lines) to the RBS random spectra using the SIMNRA code [27,28] of PLM-treated Te-hyperdoped Si samples. The inset shows Te signal for all samples where the sample names refer to the peak Te concentration.

The electrical properties of Te-hyperdoped Si samples were examined using a commercial Lakeshore Hall System (9700A) in van-der-Pauw-geometry [29]. Samples were measured in the temperature range from 2 to 300 K and a magnetic field perpendicular to the sample plane swept from -5 T to 5 T. Prior to the electrical measurements, the native $SiO_2$ layer was removed by HF etching. Subsequently, gold electrodes were sputtered onto the four corners of the square-like samples to ensure Ohmic contact [30]. Silver glue was used to contact the wires to the gold electrodes. The carrier concentrations of Te-hyperdoped Si samples obtained from the Hall effect measurements are listed in Table I, where the electrical activation efficiency of 10%-40% can be deduced. In our previous work, we have found there is a considerable amount of interstitial Te in samples with low doping concentrations. This explains the lower activation fraction for samples Te-0.25%, Te-0.5% and Te-1.0%. However, with increasing impurity concentration Te dopants tend to form substitutional dimers at the cost of interstitials, which contribute free electrons and lead to higher activation fraction.

TABLE I. Sample description and notations used in the manuscript. For the experimental samples, the depth distribution of tellurium is calculated using the SRIM code [25] and verified by RBS measurements [17]. For the doped layer, a thickness of 120 nm and a nominal tellurium peak concentration are obtained. The carrier concentrations are calculated from Hall measurements by taking an effective thickness of 120 nm.

| Computation | | | Experimental samples | | | |
|---|---|---|---|---|---|---|
| Sample ID | Tellurium concentration ($N$) (%) | Tellurium concentration ($N$) (cm$^{-3}$) | Sample ID | Nominal tellurium peak concentration ($N$) (cm$^{-3}$) | Nominal tellurium peak concentration ($N$) (%) | Measured carrier concentration at 300 K ($n$) (cm$^{-3}$) |
| Te-0.39% | 0.39 | $1.95 \times 10^{20}$ | Te-0.25% | $1.25 \times 10^{20}$ | 0.25 | $2.0 \times 10^{19}$ |
| Te-0.92% | 0.92 | $4.80 \times 10^{20}$ | Te-0.50% | $2.50 \times 10^{20}$ | 0.50 | $8.5 \times 10^{19}$ |
| Te-1.56% | 1.56 | $7.80 \times 10^{20}$ | Te-1.00% | $5.00 \times 10^{20}$ | 1.00 | $1.7 \times 10^{20}$ |
| | | | Te-1.50% | $7.50 \times 10^{20}$ | 1.50 | $4.4 \times 10^{20}$ |
| | | | Te-2.00% | $1.00 \times 10^{21}$ | 2.00 | $6.0 \times 10^{20}$ |
| | | | Te-2.50% | $1.25 \times 10^{21}$ | 2.50 | $8.3 \times 10^{20}$ |

## III. RESULTS

## A. Band structure and DOS

In the band structure and density of states (DOS) calculations of Te-hyperdoped Si we considered the Te substitutional impurities, namely substitutional single Te (Te$_{Si}$) and substitutional Te dimers (Te$_{Si}$-Te$_{Si}$), which represent the large majority of defect type present in hyperdoped Si. At the Te concentrations considered in the present study, the interstitial Te impurities exhibit a significantly higher formation energy [24]. Particularly, according to the previous study in Ref. [24], Te$_{Si}$-Te$_{Si}$ has the lowest formation energy among all types of defects considered and becomes the dominant configuration as effective donors with increasing Te concentration, especially in the metallic regime [24]. The computed electronic DOS and the electronic band structure for Te$_{Si}$ and Te$_{Si}$-Te$_{Si}$ in Te-hyperdoped Si at three different Te doping concentrations ($x = 0.39\%$, 0.92% and 1.56%) are displayed in FIG. 2 and FIG. 3. The doping concentration range was chosen to basically cover the transition from the insulating to the metallic regime. In this section (III. A) we present the first-principles simulations of electronic states, and in the next section (III. B) the electrical conductivity measurements. The random distribution of dopants lifts the translational invariant symmetry; thus, for a direct comparison of simulated electronic states with experimental data the DOS becomes the relevant quantity rather than the band-structure. However, it can be convenient to study the evolution of the electronic states produced by doping, looking to the modification of the IB in the band-structure obtained by the super-cell method.

The calculated band structure and the DOS of Te-hyperdoped Si shown in FIG. 2 and FIG. 3 demonstrate the modification of the electronic properties and, in particular, the evolution of the IB which at low concentration is in the indirect band gap originating from the conduction band minimum (CBM) and the valence band maximum (VBM) of pure Si. As the concentration of Te is increased, the local CBM located at the Γ-point is pushed downwards due to the interactions between Te and Si. As displayed in FIG. 2(a) and (b), for Te$_{Si}$ single impurities at $x = 0.39\%$, the IB is very close to the bottom of the conduction band, thus forming a semi-metallic system with a very small or vanishing gap. Since IMT is usually associated with the merging of the impurity band with the conduction band [14], the computed DOS denotes that for a system composed only of Te$_{Si}$ single impurities, the value $x = 0.39\%$ is very close to the critical concentration at which the IMT occurs. At variance, as displayed in the insert of FIG. 2(c) and (d), for Si doped with Te$_{Si}$ dimers at $x = 0.39\%$, the IB is still separated from the CBM (by approximately 0.04 eV). The small gap between the CBM and the completely filled IB leads to a vanishing contribution to the conduction at lowest temperatures, since there are no nearby empty states. Therefore, at this concentration Te dimer hyperdoped Si is in the insulating state and exhibits only thermally activated conductivity, being qualitatively consistent with previous experimental investigations [17,31,32].

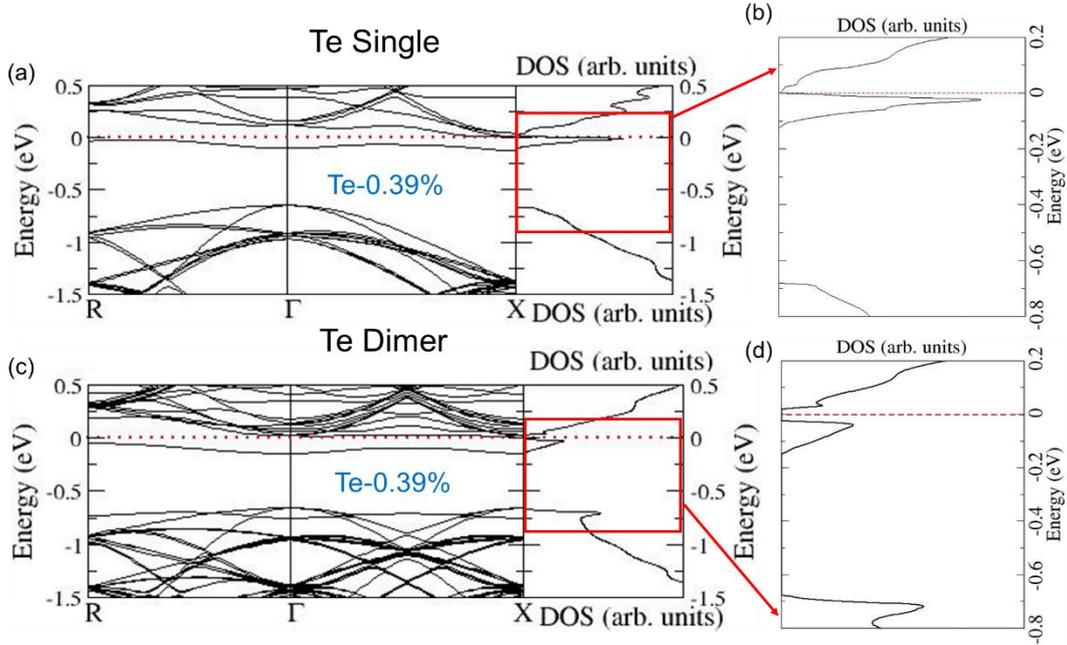

FIG. 2 *Ab-initio* calculations of the electronic band structure (along high symmetry directions of the cubic Brillouin zone) and the corresponding electronic density of states (DOS) for single Te substitutional (Te$_{Si}$) dopants ((a)) and substitutional Te dimers (Te$_{Si}$-Te$_{Si}$) ((c)) in Te-hyperdoped Si at Te concentration of $x$ = 0.39 %. (b) Zoom out of the selected area in (a). (d) Zoom out of the selected area in (c).

At the concentration of $x$ = 0.39%, the IBs corresponding to Te$_{Si}$ and Te$_{Si}$-Te$_{Si}$ are relatively flat. As the Te concentration increases, the impurity band is getting broader (see FIG. 3(a)-(d)), which indicates the increased delocalization of the impurity states. This is due to the decreased spacing between impurities, which results in the increased dispersion and interactions between neighboring Te impurities [17]. Particularly, as shown in FIG. 3(a) and FIG. 3(b), at the doping concentration $x$ = 0.92%, the IB is widened and tends to partially overlap with the CB. This band overlap produces the full hybridization of IB of higher energy with the conduction states of lower energy. At the same time the Fermi level is no longer located in the region of the forbidden energy as for the insulating state, but gradually enters into the CB, producing a metallic state. Thus, the low-lying conduction-band-like states are available for the charge transport without thermal activation, implying the occurrence of the IMT upon doping. At Te concentrations of 1.56%, the IB is further widened to 0.52 eV for Te$_{Si}$ and 0.40 eV for Te$_{Si}$-Te$_{Si}$ in Te-hyperdoped Si, respectively (see FIG. 3(c) and (d)). Particularly, for the Te$_{Si}$-Te$_{Si}$ case there is a strong hybridization of IB with states originated from Si valence bands (more details can be found in FIG 4(c) in Ref. [24]). In FIG. 3, the intersection of Fermi energy level with the conduction band is a fingerprint of the overlap of higher energy IB with lower energy Si conduction bands, as evident by looking to the DOS (while this is less evident by looking to the electronic bands that are displayed along two high symmetry directions: they account only partially the electronic structure of the whole Brillouin zone. The latter information is contained in the DOS). The band structures and DOS analysis demonstrate the delocalization of the impurity states and the eventual merging of the IB and the CB as the doping concentration increases.

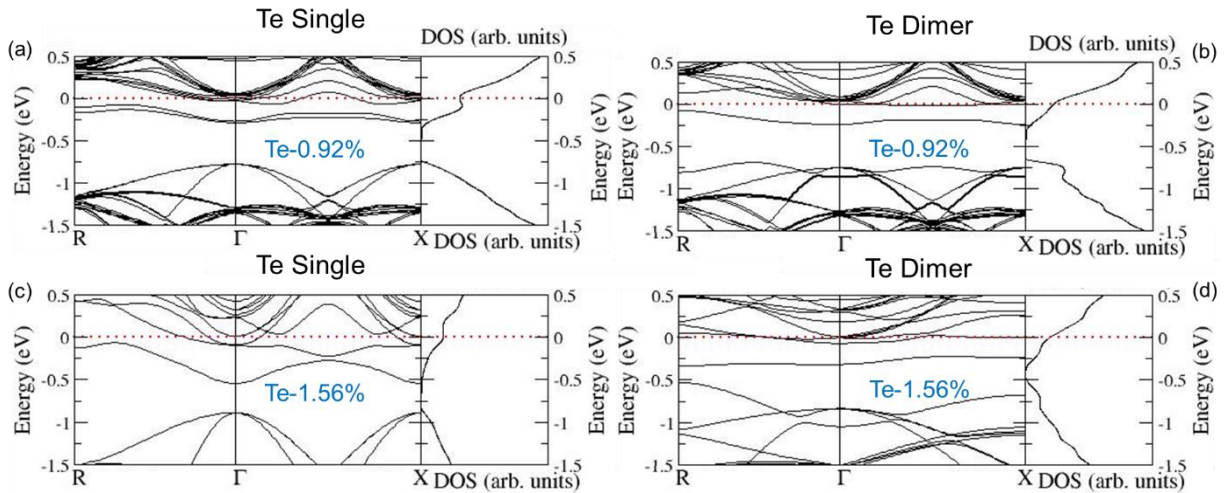

FIG. 3 *Ab-initio* calculations of the electronic band structure (along high symmetry directions of the cubic Brillouin zone) and the corresponding electronic density of states (DOS) for single Te substitutional (Te$_{Si}$) dopants ((a) and (c)) and substitutional Te dimers (Te$_{Si}$-Te$_{Si}$) ((b) and (d)) in Te-hyperdoped Si at different Te concentrations. (a) and (b), $x = 0.92$ % (top panel), (c) and (d), $x = 1.56$ % (bottom panel). The DOS for $x = 1.56$ % form Ref. [24] is shown for completeness. The zero of the energy scales (red dotted line) corresponds to the Fermi energy. Notice that the band-structure is computed only along some high symmetry directions (and thus only some electronic states are displayed), while the DOS is computed over the whole Brillouin zone (and thus all electronic states are displayed).

As displayed in FIG. 3(a)-(d), the Fermi level is located in the CB for both systems at Te doping concentration of 0.92% and 1.56%, which corresponds to the metallic state and is consistent with transport measurement results (see FIG. 4). However, for the case of Te$_{Si}$-Te$_{Si}$ at $x = 0.92$%, the Fermi energy is located very close to the bottom of the conduction band, suggesting that this value of the concentration is only slightly higher than the critical concentration of IMT, $n_{crit}$, predicted for Te$_{Si}$-hyperdoped Si. From our first principles data, by considering the ideal case in which only one type of defect is present in hyperdoped Si, we can argue that the IMT for isolated Te$_{Si}$ single system occurs at a lower Te concentration than for Te$_{Si}$-Te$_{Si}$ dimer. In particular, by the comparison between the electronic states obtained by first principles data and the experimental conductivity presented in section III. B, within the approximations used, our first principles results for Te$_{Si}$-Te$_{Si}$ dimer are in agreement with experimental data qualitatively, thus suggesting that very probably the Te$_{Si}$-Te$_{Si}$ plays a fundamental role as the driving force of IMT.

## B. Temperature dependence of the conductivity

It is known that the real difference between insulators and metals is revealed only at zero temperature. The metallic state is defined by exhibiting finite conductivity as the temperature (T) approaches zero, whereas insulators exhibit vanishing conductivity as T approaches zero. The temperature-dependent conductivity is shown in FIG. 4(a). The sample with the lowest Te concentration (Te-0.25%) clearly tends towards vanishing conductivity, the same is concluded for Te-0.50%, although measurements at lower temperature would be desirable to make it clearer. Samples with the highest Te concentrations (Te-2.0% and Te-2.5%) exhibit a much higher conductivity which is also insensitive to temperature down to 2 K. Note that their conductivities are comparable to that of shallow-impurity-doped Si with just-metallic concentrations [5] and higher than those of metallic Si doped with S and Se [13,14]. Therefore, these two samples are metallic, whereas samples Te-0.25% and Te-0.50% are insulating. The samples Te-1.0% and Te-1.5% lie near the critical regime of IMT which will be analyzed further in the following.

Figure 4(b) displays a rigorous experimental evidence of an IMT in the PLM-treated Te-hyperdoped Si samples, which exhibit peak Te concentrations ($N$) from $1.25 \times 10^{20}$ cm$^{-3}$ to $1.25 \times 10^{21}$ cm$^{-3}$ and carrier concentration ($n$) from $2.0 \times 10^{19}$ cm$^{-3}$ to $8.3 \times 10^{20}$ cm$^{-3}$ at 300 K. We compare the carrier concentrations measured at 2 K and 300 K, which are calculated by taking an effective thickness of 120 nm. For sample Te-0.25% and sample Te-0.50%, the carrier concentration at 2 K is substantially lower than that at 300 K, a clear evidence for electrical freeze-out, i.e. the donor electrons return into the localized ground states from thermally excited states as the temperature decreases. However, samples with Te peak concentration higher than $5.7 \times 10^{20}$ cm$^{-3}$ (Te-1.0%) exhibit temperature-independent carrier concentrations. Here, sample Te-1.0% ($n = 1.7 \times 10^{20}$ cm$^{-3}$) seems to be right at the border, suggesting that Te-1.5% is already metallic. The critical Te concentration for IMT is slightly higher than the value calculated by first-principles calculations shown in III.A. This can be understood by the fact that the samples contain defects and some Te impurities are not in substitutional positions.

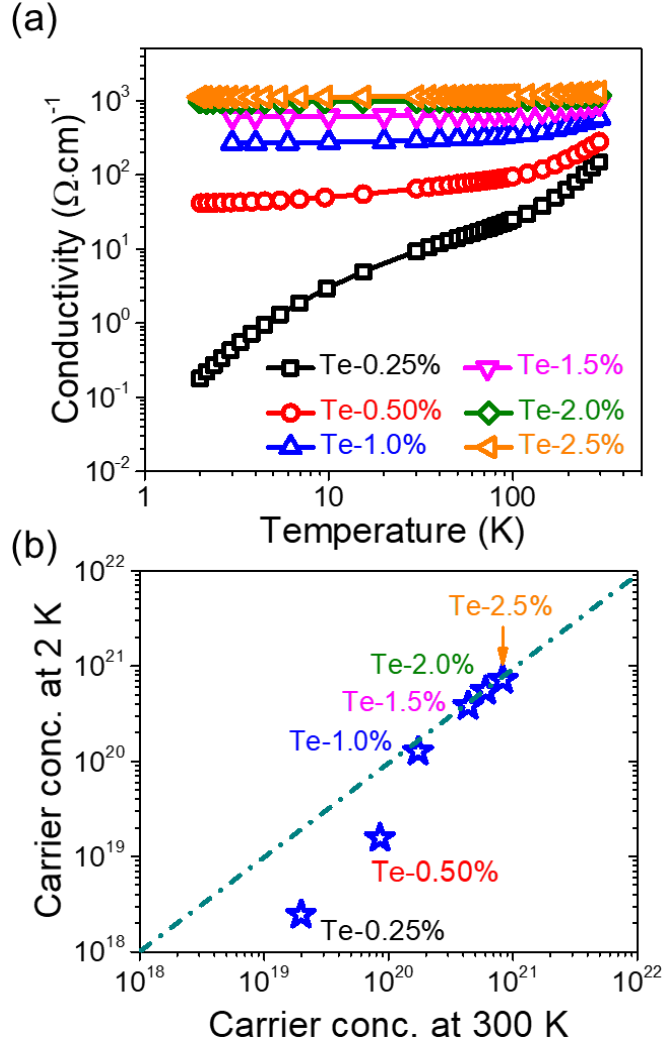

FIG. 4. Electrical properties for PLM-treated Te-hyperdoped Si samples with different Te concentrations. (a) The temperature-dependent conductivity of the Te-hyperdoped Si samples. (b) carrier concentration measured at 2 K vs. that at 300 K. The dashed line shows the metallic behavior. Samples with Te concentration higher than 1.0 % show metallic behavior, while sample Te-0.25% and sample Te-0.50% exhibit carrier freeze-out and behave as insulators.

## IV. DISCUSSION

### A. Critical impurity concentration of IMT

We first discuss the critical impurity concentration of IMT obtained from the transport experiments and the Mott theoretical calculation outcome compared to the first principles computational results. As confirmed in FIG. 4, samples Te-1.5% and in particular Te-1.0% appear to be in the transition regime. Their carrier concentrations are $4.4 \times 10^{20}$ cm$^{-3}$ and $2 \times 10^{20}$ cm$^{-3}$, respectively, which corresponds to an activation efficiency around 20%-30%. In addition, as Mott originally derived, the IMT in the group-IV semiconductors can be estimated as:

$$n_{crit}^{1/3} a_H = 0.25 \quad (1)$$

where $a_H$ is the effective Bohr radius of the donor electrons, [1,33]:

$$a_H = \frac{e^2}{8\pi\varepsilon_0 \varepsilon_r E} \quad (2)$$

where $E$ is the binding (or activation) energy of the localized states, $\varepsilon_0$ and $\varepsilon_r$ are the permittivity of free space and the high-frequency dielectric constant, respectively. Thus, by taking into account the binding energy as 199 meV [31,34], the isotropic Bohr radius is calculated as 3.1 Å and the critical carrier concentration of Te hyperdoped Si is approximately $n_{crit} = 5.24 \times 10^{20}$ cm$^{-3}$. However, in the Mott criterion, the $a_H$ of an isolated center is defined as an appropriate radius associated with a realistic wave function for the localized state in the low-electron-density regime. In this case, the broadening of the electron wavefunction increases with the doping concentration. Therefore, the critical carrier concentration obtained from the experimental data (between $1.7 \times 10^{20}$ cm$^{-3}$ and $4.4 \times 10^{20}$ cm$^{-3}$, see the conclusion from FIG. 4) is actually lower than that from the value computed by the Mott criterion.

## B. Critical behavior of temperature-dependent transport

In this section, the underlying physics behind the electrical properties of all the Te-hyperdoped Si samples will be explored by modelling the experimental data.

### *1. Metallic samples*

The conductivity in the metallic phase can be modelled to the form [35]:

$$\sigma(T) = \sigma_0 + mT^s \quad (3)$$

where $\sigma_0$ represents the zero-Kelvin conductivity, $m$ is a constant and the temperature exponent $s$ is related to the scattering mechanism in the metallic phase. Electron-electron interactions in disordered systems lead to the lowest-order correction $mT^{1/2}$ to $\sigma_0$ [5, 36], therefore we fixed the temperature exponent $s$ as 1/2 in the modelling. As Mott originally proposed, a minimum metallic conductivity ($\sigma_M$) at $T = 0$ K can be defined as [37]

$$\sigma_M = \frac{Ce^2}{\hbar d_c} \quad (4)$$

where the numerical coefficient $C \approx 0.12$ is for $n$-Si [38] and $d_c$ is the average spacing between impurity atoms at the critical concentration ($n_c$). Here $\sigma_M$ with the value of 247 ($\Omega$cm)$^{-1}$ is obtained by using $d_c = (n_{crit})^{-1/3}$. FIG. 5(a) shows the plot $\sigma$ (conductivity) vs $T^{1/2}$ for samples with high Te concentration $N$. The relationship, $\sigma(T) = \sigma_0 + mT^{1/2}$, is approximately obeyed between 2 K and 25 K, furthermore, the overall $T$ dependence for the samples with different Te concentrations $N$ makes the extrapolation to $T = 0$ rather unambiguous. Samples with high Te concentrations exhibit $\sigma_0 > \sigma_M$, indicative of the metallic phase and consistent with the Mott picture. FIG. 5(b) displays the extrapolated $\sigma_0$ as a function of carrier concentration $n$. The critical behaviour is highly suggested by the sharpness of the transition, which is remarkable and is qualitatively close to the discontinuity predicted by Mott [37]. The fitting of the equation (3) yields $\sigma_0(n, 0) = \sigma_0 \; (\frac{n}{n_c} - 1)^\mu$ with $\sigma_0 = 985$ $(\Omega cm)^{-1}$ (almost four times of $\sigma_M$), $n_c = 1.54 \times 10^{20}$ cm$^{-3}$ and the critical conductivity exponent $\mu = 0.48 \pm 0.07$.

FIG. 5(c) shows temperature correction ($m$) (extrapolated from FIG. 5(a)) plotted against carrier concentration $n$ for samples with carrier concentration in the range of $1.1 < n/n_c < 5.4$. As being well established, two classical models were proposed for the explanation of $m$ (the correction to $\sigma_0$): the scaling theory of localization [2,39] and the Coulomb interaction with electron-electron scattering [40-42]. The latter is valid for $k_F l \gg 1$ [5]. Here $k_F$ is the Fermi wavevector given by [43]:

$$k_F = (\frac{3\pi^2}{g} n)^{1/3} \quad (5)$$

where $g = 6$ is the number of equivalent minima in the conduction band of Si; $l$ is the mean free path, which scales as:

$$l = \frac{3\pi^2}{4} \frac{\hbar}{e^2} \frac{\sigma_0}{k_F^2} \quad (6)$$

As shown in FIG. 5(c), a positive value of $m$ term is produced for insulating samples. This would be consistent with the scaling theory of localization extended to include inelastic scattering but neglecting Coulomb interactions [2,39]. Also, here $k_F l$ is no longer significantly greater than 1. $m$ does not significantly change in the critical transition regime ($n/n_c \approx 1.1$~2.9), while towards larger $n$, $m$ changes the sign from positive to negative for samples with $n/n_c > 2.9$. This is in agreement with the reported work (P or As doped Si) [5,44,45], resulting from the Coulomb interactions with electron-electron scattering in the presence of random impurities. However, one has to note that the lowest measurement temperature is 2 K in our case, which is much higher than some mK used in other cases. The lack of the data at mK temperatures and the detailed variation of Te concentration around 1.0% could result in errors in both $n_c$ and $m$.

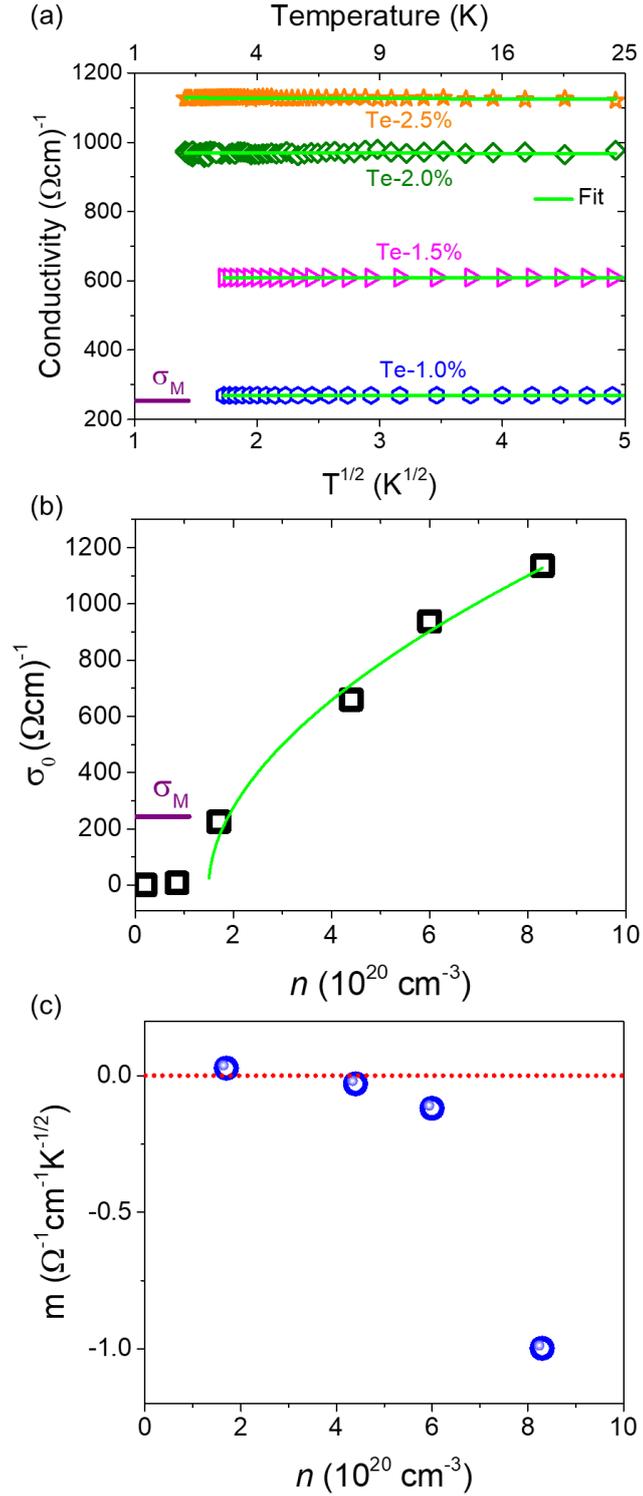

FIG. 5. Analysis of temperature-dependent transport properties for metallic samples. (a) Electrical conductivity $\sigma$ vs $T^{-1/2}$ for Te-hyperdoped Si samples with high Te concentration $N$. (b) Extrapolated conductivity $\sigma_0$ as a function of carrier concentration $n$. The $\sigma_0$ data were fitted by $\sigma_0(n,0) = \sigma_0 \left(\frac{n}{n_c} - 1\right)^\mu$ with $\mu = 0.48 \pm 0.07$, $n_c = 1.54 \times 10^{20}\,\text{cm}^{-3}$. (c) Coefficient $m$ of the $T$ dependence of $\sigma$ vs. carrier concentration.

## 2. Insulating samples

The electrical conductance in the insulating phase at low temperature can be achieved by hopping through the localized deep levels [46]. The conductivity scales as:

$$\sigma(T) = \sigma_0 exp[-(\frac{T_p}{T})^p] \qquad (7)$$

The pre-factor $\sigma_0$ and the characteristic temperature $T_p$ are related to material parameters by different relationships for each value of *p*. The exponent *p* depends on the temperature and the shape of the density of the states (DOS) near the Fermi level. In detail, $p$ = 1/4, 1/3, 1/2 or 1 corresponds to Mott-law variable-range hopping for 3D and 2D systems, the Efros-Shklovskii-type variable-range hopping (a Coulomb gap in the DOS), and the nearest-neighbour hopping, respectively [46]. To explicitly describe the temperature dependence of the conductivity and determine the value of *p*, Zabrodskii and Zinovevawas [47] introduced the reduced activation energy *W* defined as:

$$W(T) = \frac{dln(\sigma)}{dln(T)} = \left(\frac{T}{\sigma}\right)\frac{d\sigma}{dT} \qquad (8)$$

which enables to determine which charge transport mechanism is mostly dominant among metallic, insulating, and the boundary of the IMT. Therefore, for the materials in the insulating phase, inserting equation (7) into (8) gives

$$W(T) = p\left(\frac{T_0}{T}\right)^p \qquad (9)$$

As shown in FIG. 6(a), the data is replotted as $W(T) = \frac{dln(\sigma)}{dln(T)}$ versus temperature on a log-log scale. For conductivity described as in Eq. (9), the slope of log*W* versus log*T* yields the value of $p$ = 0.45 ± 0.04 in the temperature range of *T* < 50 K. By this analysis, the values are very close to $p$ = 1/2, which corresponds to the conduction of the Efros-Shklovskii-type variable-range hopping (ES law) [46]. This is the case where the impurity levels are deep enough and in turn the Coulomb gap is fairly large and has a certain vicinity of the Fermi level. The experimental results here indicate that the Coulomb gap is symmetric with respect to the Fermi level in the IB. This is consistent with the DFT calculation results considering the fact that Te$_{Si}$ and Te$_{Si}$-Te$_{Si}$ co-exist in sample Te-0.25% [24].

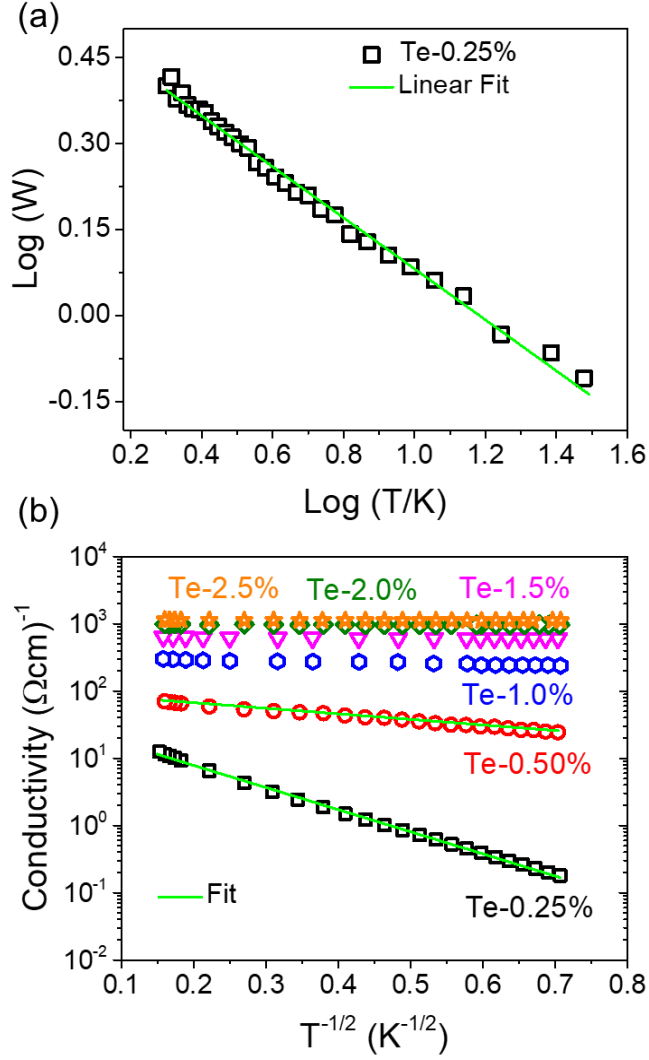

FIG. 6. Analysis of temperature-dependent transport properties for semi-insulating samples. (a) $W(T)$ versus $T$ on a log scale. The green solid line shows $p = 0.45 \pm 0.04$. (b) Conductivity of the Te-hyperdoped Si samples as a function of $T^{-1/2}$. The solid lines are fits of the experimental data by equation (7) with $p = 1/2$.

FIG. 6(b) displays the conductivity of semi-insulating samples as the function of $T^{-1/2}$. As the doping concentration approaches $n_c$, a sharp increase of conductivity is observed in the samples with small difference of impurity concentration. The fitting of variable-range hopping conductivity of insulating samples is presented in FIG. 6(b) as the solid lines. The fitting parameter $p = 1/2$ provides a reasonable fit with an average relative mean square error of 1.2% and the fitting range is restricted to T < 50 K. According to the law of Efros and Shklovskii [2], the characteristic parameter $T_P$ (in equation (7)) is related with fundamental material properties:

$$T_P = \frac{Ce^2}{4\pi\varepsilon_0\varepsilon_r k_B \xi} \quad (10)$$

where $e$ is the electron charge, $\varepsilon$ is the material permittivity, $k_B$ is the Boltzmann constant, $\xi$ is the electron correlation length and C=2.8 is a numerical coefficient [2,48,49]. $T_P$ can be obtained from the data fitting using equation (7): $T_P$ = 548 K (47 meV) for sample Te-0.25% and $T_P$ = 180 K (16 meV) for sample Te-0.50%. Knowing $T_P$, the material permittivity [50] and using equation (10), the electron correlation length $\xi$ can be computed. The electron correlation length $\xi$ increases from 7 nm in sample Te-0.25% to 22 nm in sample Te-0.50%. $\xi$ increases as the Te doping concentration increases. This indicates that the dopant concentration approaches the critical transition concentration of the IMT [49,51]. This is also corroborated in our DFT calculations.

## V. CONCLUSION

In conclusion, we have investigated the transport properties of Te-hyperdoped Si samples prepared by ion implantation followed by nanosecond pulsed-laser melting. An insulator-to-metal transition driven by increasing Te concentration is confirmed and illustrates an agreement with the DFT computational results as well as with Mott's theoretical picture. By performing the physical modelling for the temperature-dependent transport data, we have demonstrated that at sufficiently low temperatures the metallic samples show a power law dependence whereas the insulating samples reveal a variable-range-hopping type-conduction with a Coulomb gap at the Fermi level. These experimental findings have allowed us to identify the critical behavior near the IMT in Te-hyperdoped Si and have confirmed the effect of disorder and electron-electron interactions induced by Te dopants on the electron wave functions.

## ACKNOWLEDGMENTS

Authors acknowledge the ion implantation group at HZDR for performing the Te implantations. Additionally, support by the Structural Characterization Facilities at Ion Beam Center (IBC) is gratefully acknowledged. This work is funded by the Helmholtz-Gemeinschaft Deutscher Forschungszentren (HGF-VH-NG-713). The authors thank Tian Ma and Pengyu Shi for Matlab introduction. M.W. thanks financial support by Chinese Scholarship Council (File No. 201506240060). A.D. acknowledges CINECA for computational resources allocated under the ISCRA initiative (IMeCS project) and thanks R. Colnaghi for technical support on computer hardware.